\documentclass[12pt]{article}
\usepackage{amsmath}
\usepackage{amssymb}

\numberwithin{equation}{section}

\title{
HIGHER-ORDER SUSY, EXACTLY SOLVABLE POTENTIALS, AND EXCEPTIONAL ORTHOGONAL POLYNOMIALS}
\author{C. QUESNE\\
{\small \sl Physique Nucl\'eaire Th\'eorique et Physique Math\'ematique,}\\ 
{\small \sl Universit\'e Libre de Bruxelles, Campus de la Plaine CP229,} \\ 
{\small \sl Boulevard~du Triomphe, B-1050 Brussels, Belgium} \\
{\small \sl cquesne@ulb.ac.be}}
\date{ }
\begin{document}
\baselineskip=22pt plus 1pt minus 1pt
\maketitle

\begin{abstract} 
Exactly solvable rationally-extended radial oscillator potentials, whose wavefunctions can be expressed in terms of Laguerre-type exceptional orthogonal polynomials, are constructed in the framework of $k$th-order supersymmetric quantum mechanics, with special emphasis on $k=2$. It is shown that for $\mu=1$, 2, and 3, there exist exactly $\mu$ distinct potentials of $\mu$th type and associated families of exceptional orthogonal polynomials, where $\mu$ denotes the degree of the polynomial $g_{\mu}$ arising in the denominator of the potentials. 
\end{abstract}

\noindent
Running head: Higher-order SUSY


%
%
\newpage
\section{Introduction}

In recent years, one of the most interesting developments in quantum mechanics has been the construction of new exactly solvable potentials connected with the appearance of families of exceptional orthogonal polynomials (EOP) in mathematical physics. In contrast with families of classical orthogonal polynomials, which start with a constant, the exceptional ones start with some polynomial of degree $m \ge 1$, but still form complete sets with respect to some positive-definite measure.\par
%
%
The first examples of EOP, the so-called Laguerre- and Jacobi-type $X_1$ families, corresponding to $m=1$, were proposed in the context of Sturm-Liouville theory \cite{gomez09, gomez10a}. They were then applied to quantum mechanics and proved to be related to some exactly solvable rational extensions of known potentials, with the additional interesting property of being translationally shape invariant \cite{cq08, bagchi09a} (although at that time the list of additive shape-invariant potentials was thought to be complete \cite{cooper, carinena, bougie}).\par
%
%
Shortly thereafter, the first examples of $X_2$ EOP and of related shape-invariant potentials were proposed in the framework of conventional supersymmetric quantum mechanics (SUSYQM), suggesting a growing complexity with $m$ \cite{cq09}. Two distinct families of Laguerre- and Jacobi-type $X_m$ families were then constructed for arbitrary large $m$ \cite{odake09, odake10a} and their properties were studied \cite{odake10b, ho09}. Several approaches were considered in connection with the Darboux-Crum transformation \cite{gomez10b, gomez11a, sasaki}, the Darboux-B\"acklund one \cite{grandati11a, grandati11b} or the prepotential method \cite{ho11a, ho11b}. Possible applications to position-dependent mass systems \cite{midya}, to $\cal PT$-symmetric potentials \cite{bagchi09a, bagchi10}, to conditionally-exactly solvable ones \cite{dutta, odake11a}, and to the Dirac and Fokker-Planck equations \cite{ho11c} were studied.\par
%
%
Very recently, the $X_m$ EOP were generalized to multi-indexed families by means of multi-step Darboux algebraic transformations \cite{gomez11b} or the Crum-Adler mechanism \cite{odake11b}.\par
%
%
The purpose of the present work is to show how the related exactly solvable potentials arise in an (essentially equivalent) simple extension to higher order of the SUSYQM approach used in \cite{bagchi09a, cq09}. For simplicity's sake, we shall restrict ourselves here to rational extensions of the radial oscillator connected with Laguerre-type EOP.\par 
%
%
\section{Second-order SUSYQM}

Going from first- to second-order SUSYQM may be achieved in two slightly different ways, corresponding to the parasupersymmetric (PSUSY) scheme \cite{rubakov} or to the second-derivative (SSUSY) setting \cite{andrianov, bagrov, bagchi99, aoyama, fernandez}.\par
%
%
In the former approach, one starts with a pair of first-order SUSYQM partners (in units $\hbar = 2m = 1$)
\begin{equation}
\begin{split}
  & H^{(+)} = A^{\dagger} A = - \frac{d^2}{dx^2} + V^{(+)}(x) - E, \quad H^{(-)} = A A^{\dagger} = 
       - \frac{d^2}{dx^2} + V^{(-)}(x) - E, \\
  & A^{\dagger} = - \frac{d}{dx} + W(x), \quad A = \frac{d}{dx} + W(x), \quad V^{(\pm)}(x) = W^2(x) \mp 
       W'(x) + E,  
\end{split}
\end{equation}
which intertwine with the first-order differential operators $A$ and $A^{\dagger}$ as $A H^{(+)} = H^{(-)} A$ and $A^{\dagger} H^{(-)} = H^{(+)} A^{\dagger}$. Here $W(x)$ is the superpotential, which can be expressed as $W(x) = - \phi'(x)/\phi(x)$ in terms of a (nodeless) seed solution $\phi(x)$ of the initial Schr\"odinger equation
\begin{equation}
  \left(- \frac{d^2}{dx^2} + V^{(+)}(x)\right) \phi(x) = E \phi(x),  \label{eq:SE}
\end{equation}
$E$ is the factorization energy, assumed smaller than or equal to the ground-state energy $E^{(+)}_0$ of $V^{(+)}$, and a prime denotes a derivative with respect to $x$. We shall only consider here the case where $E < E^{(+)}_0$, in which occurrence $\phi(x)$ is nonnormalizable, and we shall assume that the same holds true for $\phi^{-1}(x)$. Then $H^{(+)}$ and $H^{(-)}$ turn out to be isospectral \cite{cooper}.\par
%
%
Next, we consider a second pair of first-order SUSYQM partners $\tilde{H}^{(+)}$ and $\tilde{H}^{(-)}$ with the same characteristics and distinguish all related quantities from those pertaining to the first pair by tildes. If we choose $\tilde{V}^{(+)}(x)$ such that $\tilde{V}^{(+)}(x) = V^{(-)}(x)$, then both pairs of SUSYQM partners $(H^{(+)}, H^{(-)})$ and $(\tilde{H}^{(+)}, \tilde{H}^{(-)})$ can be glued together, so that we get a second-order PSUSY system or, equivalently, a reducible SSUSY one.\par
%
%
The latter is described in terms of two Hamiltonians
\begin{equation}
  h^{(1)} = - \frac{d^2}{dx^2} + V^{(1)}(x), \qquad h^{(2)} = - \frac{d^2}{dx^2} + V^{(2)}(x), 
\end{equation}
which intertwine with some second-order differential operators
\begin{equation}
  {\cal A}^{\dagger} = \frac{d^2}{dx^2} - 2p(x) \frac{d}{dx} + q(x), \qquad {\cal A} = \frac{d^2}{dx^2} + 2p(x)
  \frac{d}{dx} + 2p'(x) + q(x),
\end{equation}
as ${\cal A} h^{(1)} = h^{(2)} {\cal A}$ and ${\cal A}^{\dagger} h^{(2)} = h^{(1)} {\cal A}^{\dagger}$, so that the functions $p(x)$, $q(x)$ and the potentials $V^{(1,2)}(x)$ are constrained by the relations
\begin{equation}
\begin{split}
  & q(x) = - p' + p^2 - \frac{p''}{2p} + \left(\frac{p'}{2p}\right)^2 - \frac{c^2}{16p^2},  \\
  & V^{(1,2)}(x) = \mp 2p' + p^2 + \frac{p''}{2p} - \left(\frac{p'}{2p}\right)^2 + \frac{c^2}{16p^2}, 
\end{split}  \label{eq:SSUSY}
\end{equation}
where $c$ is some integration constant.\par
%
%
The relation between both approaches follows from the equations $h^{(1)} = H^{(+)} + \frac{c}{2}$, $h^{(2)} = \tilde{H}^{(-)} - \frac{c}{2}$, ${\cal A}^{\dagger} = A^{\dagger} \tilde{A}^{\dagger}$, and ${\cal A} = \tilde{A} A$. It turns out that $h^{(1)}$ and $h^{(2)}$ are both partners of some intermediate Hamiltonian $h = H^{(-)} + \frac{c}{2} = \tilde{H}^{(+)} - \frac{c}{2}$, that the constant $c$ is related to the two factorization energies through $c = E - \tilde{E}$, and that the function $p(x)$ can be expressed in terms of the two superpotentials as $p(x) = \frac{1}{2} (W + \tilde{W})$.\par
%
%
Instead of $\phi$ and $\tilde{\phi}$, we may start from two seed solutions $\phi_1$ and $\phi_2$ of the initial Schr\"odinger equation (\ref{eq:SE}) with respective energies $E_1$ and $E_2$ (less than $E^{(+)}_0$) and such that $\phi_1^{-1}$ and $\phi_2^{-1}$ are nonnormalizable. Then, on choosing $\phi = \phi_1$ and $\tilde{\phi} = A \phi_2 = {\cal W}(\phi_1, \phi_2)/\phi_1$, so that $E = E_1$ and $\tilde{E} = E_2$, we get
\begin{equation}
  p(x) = - \frac{{\cal W}'(\phi_1, \phi_2)}{2 {\cal W}(\phi_1, \phi_2)} = - \frac{(E_1 - E_2) \phi_1 \phi_2}
 {2 {\cal W}(\phi_1, \phi_2)}, 
\end{equation}
where ${\cal W}(\phi_1, \phi_2)$ denotes the Wronskian of $\phi_1(x)$ and $\phi_2(x)$. Since, from (\ref{eq:SSUSY}), the SSUSY partner potential can be written as $V^{(2)}(x) = V^{(1)}(x) + 4p'(x)$, it is clear that it can be completely determined from the knowledge of this Wronskian.\par
%
%
\section{Radial Harmonic Oscillator in Second-order SUSYQM}

Let us consider a radial oscillator potential
\begin{equation}
  V_l(x) = \frac{1}{4} \omega^2 x^2 + \frac{l(l+1)}{x^2},  \label{eq:HO}
\end{equation}
where $\omega$ and $l$ denote the oscillator frequency and the angular momentum quantum number, respectively, and the range of $x$ is the half-line $0 < x < \infty$. As well known, the corresponding Schr\"odinger equation has an infinite number of bound-state wavefunctions, which, up to some normalization factor, can be written as
\begin{equation}
  \psi^{(l)}_{\nu} \propto x^{l+1} e^{- \frac{1}{4} \omega x^2} L^{(l + \frac{1}{2})}_{\nu}(\tfrac{1}{2} \omega
  x^2) \propto \eta_l(z) L^{(\alpha)}_{\nu}(z), \qquad \nu = 0, 1, 2, \ldots,
\end{equation}
with
\begin{equation}
  z = \tfrac{1}{2} \omega x^2, \qquad \alpha = l + \tfrac{1}{2}, \qquad \eta_l(z) = z^{\frac{1}{4}(2\alpha+1)}
  e^{- \frac{1}{2}z}, \label{eq:z-alpha}
\end{equation}
and $L^{(\alpha)}_{\nu}(z)$ some Laguerre polynomial. The associated bound-state energies are given by
\begin{equation}
  E^{(l)}_{\nu} = \omega (2\nu + l + \tfrac{3}{2}) = \omega (2\nu + \alpha + 1).  \label{eq:spectrum}
\end{equation}
\par
%
%
Motivated by the experience gained in the first-order SUSYQM approach (see Eqs.\ (2.12), (2.16), and (2.18) of Ref.\ 7), as well as by subsequent developments \cite{odake09, odake10a, gomez10b}, we may consider two different types of seed solutions $\phi(x)$ with properties as required in Sec.\ 2, namely
\begin{equation}
  \phi^{\rm I}_{lm}(x) = \chi^{\rm I}_l(z) L^{(\alpha)}_m(-z) \propto x^{l+1} e^{\frac{1}{4} \omega x^2} 
  L^{(l+\frac{1}{2})}_m(- \tfrac{1}{2} \omega x^2),
\end{equation}
\begin{equation}
  \phi^{\rm II}_{lm}(x) = \chi^{\rm II}_l(z) L^{(-\alpha)}_m(z) \propto x^{-l} e^{-\frac{1}{4} \omega x^2} 
  L^{(-l-\frac{1}{2})}_m(\tfrac{1}{2} \omega x^2),
\end{equation}
with
\begin{equation}
  \chi^{\rm I}_l(z) = z^{\frac{1}{4}(2\alpha+1)} e^{\frac{1}{2}z}, \qquad \chi^{\rm II}_l(z) = 
  z^{-\frac{1}{4}(2\alpha-1)} e^{-\frac{1}{2}z}, 
\end{equation}
and corresponding energies
\begin{equation}
  E^{\rm I}_{lm} = - \omega(\alpha + 2m + 1), \qquad E^{\rm II}_{lm} = - \omega(\alpha - 2m - 1),
\end{equation}
respectively. Such seed solutions are related to the two families L1 and L2 of Laguerre-type $X_m$ EOP. Note that for type II, $\alpha$ must be greater than $m$.\par
%
%
Our purpose is to construct some rationally-extended radial oscillator potentials $V_{l,\rm{ext}}(x)$ with a given $l$ by using two seed solutions $\phi_1$ and $\phi_2$, as explained in Sec.\ 2. Taking into account that the order of $\phi_1$ and $\phi_2$ is irrelevant, there are three types of possibilities for the pair $(\phi_1, \phi_2)$ and it turns out that in each case, we have to start from a potential $V^{(+)}(x) = V_{l'}(x)$ with some different $l'$. The corresponding Wronskian ${\cal W}(\phi_1(x), \phi_2(x))$ can be written in terms of some $\mu$th-degree polynomial in $z$, $g_{\mu}(z)$, itself expressible in terms of a Wronskian $\tilde{{\cal W}}(f(z), g(z))$ of some appropriate functions of $z$, as follows:
\begin{eqnarray}
  & (i) \; &  V^{(+)} = V_{l-2}, \quad \phi_1 = \phi^{\rm I}_{l-2,m_1}, \quad \phi_2 = \phi^{\rm I}_{l-2,m_2}, 
         \quad 0 \le m_1 < m_2, \nonumber \\
  & & {\cal W}(\phi_1, \phi_2) = \omega x (\chi^{\rm I}_{l-2})^2 g_{\mu}(z), \nonumber \\
  & & g_{\mu}(z) = \tilde{{\cal W}}(L^{(\alpha-2)}_{m_1}(-z), L^{(\alpha-2)}_{m_2}(-z)), \quad
     \mu = m_1 +  m_2 - 1; \label{eq:i}  \\
  & (ii) \; & V^{(+)} = V_{l+2}, \quad \phi_1 = \phi^{\rm II}_{l+2,m_1}, \quad \phi_2 = \phi^{\rm II}_{l+2,m_2}, 
         \quad 0 \le m_1 < m_2 < \alpha + 2, \nonumber \\
  & & {\cal W}(\phi_1, \phi_2) = \omega x (\chi^{\rm II}_{l+2})^2 g_{\mu}(z), \nonumber \\
  & & g_{\mu}(z) = \tilde{{\cal W}}(L^{(-\alpha-2)}_{m_1}(z), L^{(-\alpha-2)}_{m_2}(z)), 
         \quad \mu = m_1 + m_2 - 1; \label{eq:ii}  \\ 
  &  (iii) \; & V^{(+)} = V_l, \quad \phi_1 = \phi^{\rm I}_{l,m_1}, \quad \phi_2 = \phi^{\rm II}_{l,m_2}, 
        \quad 0 \le m_1, \quad 0 \le m_2 < \alpha, \nonumber \\
  & & {\cal W}(\phi_1, \phi_2) = \frac{2}{x} \chi^{\rm I}_l \chi^{\rm II}_l g_{\mu}(z), \nonumber \\
  & & g_{\mu}(z) = z \tilde{{\cal W}}(L^{(\alpha)}_{m_1}(-z), L^{(-\alpha)}_{m_2}(z)) - (z + \alpha) 
        L^{(\alpha)}_{m_1}(-z) L^{(-\alpha)}_{m_2}(z), \nonumber \\
  & &\quad \mu = m_1 + m_2 + 1.
\end{eqnarray}
\par
%
%
In all three cases, we can write (provided $g_{\mu}$ does not have any zero on the half-line)
\begin{equation}
\begin{split}
  V^{(1)} & = V_{l'} - \frac{1}{2} (E_1 + E_2), \\
  V^{(2)} & = V_l - \omega \left\{2 \frac{\dot{g}_{\mu}}{g_{\mu}} + 4z \left[\frac{\ddot{g}_{\mu}}{g_{\mu}}
          - \left(\frac{\dot{g}_{\mu}}{g_{\mu}}\right)^2\right]\right\} - \frac{1}{2} (E_1 + E_2) + C, 
\end{split}
\end{equation}
where a dot denotes a derivative with respect to $z$ and $C = - 2\omega$, $2\omega$, or 0 in case $(i)$, $(ii)$, or $(iii)$, respectively. In the associated PSUSY approach, which we cannot detail here due to space restrictions, the intermediate potential is some $V_{l-1, {\rm ext}}$, $V_{l+1, {\rm ext}}$, or $V_{l+1, {\rm ext}}$ potential. Note that it reduces to some bare radial oscillator potential in the special case where $m_1 = 0$. Furthermore, if we adopt the reverse order for the $\phi$'s, e.g. $\phi_1 = \phi^{\rm II}_{l, m_2}$, $\phi_2 = \phi^{\rm I}_{l, m_1}$ in case $(iii)$, the final potential will be the same, but the intermediate one will be different, e.g. some $V_{l-1, {\rm ext}}$ potential in case $(iii)$.\par
%
%
Both $V^{(1)}$ and $V^{(2)}$ have the same bound-state energy spectrum, given by
\begin{equation}
  E^{(1)}_{\nu l} = E^{(2)}_{\nu l} = 
    \begin{cases}  
       \omega (2\nu + 2l + m_1 + m_2 - 1) & \text{in case $(i)$}, \\
       \omega (2\nu + 2l - m_1 - m_2 + 5) & \text{in case $(ii)$}, \\
       \omega (2\nu + 2l + m_1 - m_2 + 2) & \text{in case $(iii)$},
    \end{cases}
\end{equation}
where $\nu=0$, 1, 2,~\ldots. The bound-state wavefunctions $\psi^{(2)}_{\nu}(x)$ of $V^{(2)}$ can be obtained either by acting with $\cal A$ on those of $V^{(1)}$, $\psi^{(1)}_{\nu}(x) \propto \eta_{l'}(z) L^{(\alpha')}_{\nu}(z)$, or by directly inserting the expression
\begin{equation}
  \psi^{(2)}_{\nu}(x) \propto \frac{\eta_l(z)}{g_{\mu}(z)} y_n(z), \qquad n = \mu + \nu, \qquad \nu=0, 1, 2, 
  \ldots,
\end{equation}
in the Schr\"odinger equation for $V^{(2)}(x)$. As a result, we obtain the following differential equation for $y_n(z)$,
\begin{equation}
  \left[z \frac{d^2}{dz^2} + \left(\alpha + 1 - z - 2z \frac{\dot{g}_{\mu}}{g_{\mu}}\right) \frac{d}{dz} + 
  (z - \alpha) \frac{\dot{g}_{\mu}}{g_{\mu}} + z \frac{\ddot{g}_{\mu}}{g_{\mu}}\right] y_n(z) = (\mu - n)
  y_n(z).  \label{eq:diff}
\end{equation}
The orthonormality and completeness of $\psi^{(2)}_{\nu}(x)$, $\nu=0$, 1, 2,~\ldots, on the half-line imply that for any $n = \mu + \nu$, $\nu=0$, 1, 2,~\ldots, the differential equation (\ref{eq:diff}) admits a $n$th-degree polynomial solution and that when $\nu$ runs over 0, 1, 2,~\ldots, such polynomials form an orthogonal and complete set  with respect to the positive-definite measure $z^{\alpha} e^{-z} g_{\mu}^{-2} dz$. We shall denote these polynomials by $L^{\rm I, I}_{\alpha, m_1, m_2, n}(z)$, $L^{\rm II, II}_{\alpha, m_1, m_2, n}(z)$, and $L^{\rm I, II}_{\alpha, m_1, m_2, n}(z)$ in cases $(i)$, $(ii)$, and $(iii)$, respectively.\par
%
%
It is worth noting that in cases $(i)$ and $(ii)$, the differential equation (\ref{eq:diff}) can be rewritten in a slightly different form. The definitions of $g_{\mu}(z)$ in (\ref{eq:i}) and (\ref{eq:ii}), combined with Laguerre equation, indeed lead to 
\begin{equation}
  z \ddot{g}_{\mu} =
    \begin{cases}
      2z \bar{g}_{\mu} - (\alpha + z) \dot{g}_{\mu} + \mu g_{\mu} & \text{in case $(i)$}, \\
      2z \bar{g}_{\mu} + (\alpha + z) \dot{g}_{\mu} - \mu g_{\mu} & \text{in case $(ii)$},
    \end{cases}
\end{equation}
where $\bar{g}_{\mu} = \tilde{\cal W}\bigl(\dot{L}^{(\alpha - 2)}_{m_1}(-z), \dot{L}^{(\alpha - 2)}_{m_2}(-z)\bigr)$ or $\bar{g}_{\mu} = \tilde{\cal W}\bigl(\dot{L}^{(-\alpha - 2)}_{m_1}(z), \dot{L}^{(-\alpha - 2)}_{m_2}(z)\bigr)$, respectively. As a result, we get
\begin{multline}
  \left[z \frac{d^2}{dz^2} + \left(\alpha + 1 - z - 2z \frac{\dot{g}_{\mu}}{g_{\mu}}\right) \frac{d}{dz} - 
       2 \alpha\frac{\dot{g}_{\mu}}{g_{\mu}} + 2z \frac{\bar{g}_{\mu}}{g_{\mu}}\right] 
       L^{\rm I, I}_{\alpha, m_1, m_2, n}(z) \\
  = - n L^{\rm I, I}_{\alpha, m_1, m_2, n}(z)
\end{multline}
and
\begin{multline}
  \left[z \frac{d^2}{dz^2} + \left(\alpha + 1 - z - 2z \frac{\dot{g}_{\mu}}{g_{\mu}}\right) \frac{d}{dz} +
      \frac{2z}{g_{\mu}} (\dot{g}_{\mu} + \bar{g}_{\mu})\right] L^{\rm II, II}_{\alpha, m_1, m_2, n}(z) \\
  = (2\mu - n) L^{\rm II, II}_{\alpha, m_1, m_2, n}(z).
\end{multline}
The latter equation coincides with that obtained for $\hat{L}^{(\alpha, m_1, m_2)}_n(z)$ in Ref.\ 24.\par
%
%
\section{Some Simple Examples}

In cases $(i)$ and $(ii)$, the lowest-degree example for $g_{\mu}$ corresponds to $m_1=0$, $m_2=1$, leading to $g_0 = 1$ or $g_0 = -1$ and giving back Laguerre polynomials: $L^{\rm I,I}_{\alpha,0,1,n} = L^{\rm II,II}_{\alpha,0,1,n} = L^{(\alpha)}_n$.\footnote{These equalities and the following ones are of course dependent on the normalization chosen for the EOP. We assume here that the latter can be appropriately adjusted.} On assuming $m_1=0$, $m_2=m+1$ ($m \ge 1$), we get either $g_m = L^{(\alpha-1)}_m(-z)$ or $g_m = - L^{(-\alpha-1)}_m(z)$, so that $L^{\rm I,I}_{\alpha,0,m+1,n} = L^{\rm I}_{\alpha,m,n}$ and $L^{\rm II,II}_{\alpha,0,m+1,n} = L^{\rm II}_{\alpha,m,n}$; hence the associated extended potentials coincide with those already obtained from first-order SUSYQM. The next values $m_1=1$, $m_2=2$ do not provide any new result either because $g_2 = L^{(-\alpha-1)}_2(z)$ or $g_2 = - L^{(\alpha-1)}_2(-z)$ and therefore $L^{\rm I,I}_{\alpha,1,2,n} = L^{\rm II}_{\alpha,2,n}$ and $L^{\rm II,II}_{\alpha,1,2,n} = L^{\rm I}_{\alpha,2,n}$.\footnote{It is worth observing here that the last reduction has not been noted in Ref.\ 24.} Going to $m_1=1$, $m_2=3$ at last gives rise to a new result (the same in the two cases), because $g_3 = [z^3 + 3\alpha z^2 + 3(\alpha-1)(\alpha+1) z + (\alpha-1)\alpha(\alpha+1)]/3$ differs from both $L^{(\alpha-1)}_3(-z)$ and $L^{(-\alpha-1)}_3(z)$ leading to $L^{\rm I}_{\alpha,3,n}$ and $L^{\rm II}_{\alpha,3,n}$, respectively.\par
%
%
In case $(iii)$, considering either $m_1 \ge 0$, $m_2=0$ or $m_1=0$, $m_2 \ge 0$ produces an already known result $L^{\rm I,II}_{\alpha,m_1,0,n} = L^{\rm I}_{\alpha,m_1+1,n}$ or $L^{\rm I,II}_{\alpha,0,m_2,n} = L^{\rm II}_{\alpha,m_2+1,n}$, since $g_{m_1+1} = - (m_1+1) L^{(\alpha-1)}_{m_1+1}(-z)$ or $g_{m_2+1} = (m_2+1) L^{(-\alpha-1)}_{m_2+1}(z)$ as a consequence of known relations among Laguerre polynomials. The first example not derivable from first-order SUSYQM is related to $m_1 = m_2 = 1$, but the resulting $g_3$ coincides with that obtained for cases $(i)$ and $(ii)$.\par
%
%
We conclude that SSUSY does not lead to any new extended potential of linear nor quadratic type, but gives rise to a new cubic one. Its explicit form can be obtained from $V^{(2)}$ for $\mu = 3$, which we rewrite as 
\begin{equation}
  V_{l, {\rm ext}} = V_l + V_{l, {\rm rat}}, \qquad V_{l, {\rm rat}} = - \omega \left\{2 \frac{\dot{g}_3}{g_3} + 4z 
        \left[\frac{\ddot{g}_3}{g_3} - \left(\frac{\dot{g}_3}{g_3}\right)^2\right]\right\}, \label{eq:example} 
\end{equation}
by dropping the additional constant. On combining Eq.~(\ref{eq:z-alpha}) with the expression of $g_3(z)$ given above, we get
\begin{equation}
  V_{l, {\rm rat}}(x) = \frac{N_1(x)}{D(x)} + \frac{N_2(x)}{D^2(x)},  \label{eq:V-rat}
\end{equation}
where 
\begin{equation}
\begin{split}
  & N_1(x) = 12 \omega [\omega^2 x^4 - (2l+1)^2 + 28], \\
  & N_2(x) = - 288 \omega [3 (2l+1) \omega^2 x^4 + 4 (2l-1) (2l+3) \omega x^2 \\
  &\hphantom{N_2(x) =} + (2l-1) (2l+1) (2l+3)], \\
  & D(x) = (\omega x^2 + 2l + 1)^3 - 4 (3\omega x^2 + 2l +1).
\end{split}  \label{eq:example1}
\end{equation}
This result may be compared with the two cubic-type extended potentials coming from first-order SUSYQM, which are given by Eq.~(\ref{eq:example}) with either $g_3(z) = L^{(\alpha-1)}_3(-z)$ or $g_3(z) = L^{(-\alpha-1)}_3(z)$, and which can be written in the form (\ref{eq:V-rat}) with 
\begin{equation}
\begin{split}
  & N_1(x) = 12 \omega [\omega^2 x^4 - (2l-9) (2l+5)], \\
  & N_2(x) = - 144 \omega (2l+5) [(2l+9) \omega^2 x^4 + 2 (2l+3) (2l+5) \omega x^2 \\
  &\hphantom{N_2(x) =} + (2l+1) (2l+3) (2l+5)], \\
  & D(x) = (\omega x^2 + 2l + 5)^3 - 2 (2l+5) (3\omega x^2 + 6l +11),
\end{split} \label{eq:example2}
\end{equation}
or
\begin{equation}
\begin{split}
  & N_1(x) = 12 \omega [\omega^2 x^4 - (2l-3) (2l+11)], \\
  & N_2(x) = 144 \omega (2l-3) [(2l-9) \omega^2 x^4 + 2 (2l-3) (2l-1) \omega x^2 \\
  &\hphantom{N_2(x) =} + (2l-3) (2l-1) (2l+1)], \\
  & D(x) = (\omega x^2 + 2l - 3)^3 + 2 (2l-3) (3\omega x^2 + 6l - 5),
\end{split}  \label{eq:example3}
\end{equation}
respectively. Note that $l$ is restricted to $l>0$ in (\ref{eq:example1}) and (\ref{eq:example2}), and to $l>2$ in (\ref{eq:example3}). All three extended potentials have the same spectrum (\ref{eq:spectrum}) than the conventional radial harmonic oscillator potential (\ref{eq:HO}).\par
%
%
\section{Final Comments}

What has been done in detail in second-order SUSYQM can, in principle, be generalized to higher order $k$ \cite{andrianov, bagrov, bagchi99, aoyama, fernandez}. The construction of the EOP and of the related extended potentials will then be governed by the choice of $k$ seed solutions $\phi_1(x)$, $\phi_2(x)$, \ldots, $\phi_k(x)$ of the initial Schr\"odinger equation and by their corresponding Wronskian ${\cal W}(\phi_1, \phi_2, \ldots, \phi_k)$.\par
%
%
{}For $k=3$, for instance, it is easy to see that the lowest-degree $g_{\mu}$ corresponds to $\mu=3$, which may arise in the pure cases obtained from $V^{(+)} = V_{l-3}$, $\phi_i = \phi^{\rm I}_{l-3,m_i}$ or $V^{(+)} = V_{l+3}$, $\phi_i = \phi^{\rm II}_{l+3,m_i}$, with $i=1$, 2, 3, $m_1 < m_2 < m_3$, and $\mu = m_1 + m_2 + m_3 - 3$.\footnote{Here we assume $m_1 > 0$, because $m_1=0$ would lead to $g_{\mu}$'s already found from SSUSY.} Since $g_3 = \tilde{\cal W}(L^{(\alpha-3)}_1(-z), L^{(\alpha-3)}_2(-z), L^{(\alpha-3)}_3(-z)) = - L^{(-\alpha-1)}_3(z)$ and $g_3 = \tilde{\cal W}(L^{(-\alpha-3)}_1(z), L^{(-\alpha-3)}_2(z), L^{(-\alpha-3)}_3(z)) = L^{(\alpha-1)}_3(-z)$, it turns out, however, that we shall obtain the extended potentials associated with $L^{\rm II}_{\alpha, 3,n}(z)$ and $L^{\rm I}_{\alpha,3,n}(z)$, respectively.\par
%
%
We conclude that considering $k$th-order SUSYQM with $k=1$, 2, 3,~\ldots\ leads to exactly $\mu$ distinct extended radial oscillator potentials and corresponding EOP families of $\mu$th type for $\mu=1$, 2, and 3. Whether this result  may be generalized to higher values of $\mu$ is an interesting open conjecture.\par
%
%
Other important points for future study are the construction of extended potentials of Morse or Coulomb type, also connected with Laguerre-type EOP (see, e.g., Refs.\ 16 and 18), as well as that of extended potentials related to Jacobi-type EOP. A direct proof of the shape invariance of the new potentials remains to be given. The existence of different intermediate Hamiltonians, as observed in the PSUSY approach of Sec.\ 3, is also worth analyzing along the lines of type A $\cal N$-fold supersymmetry \cite{aoyama, bagchi09b}.\par
%
%
\section*{Acknowledgments}

The author would like to thank Y.\ Grandati for several useful discussions. Some interesting comments from A.\ Khare, M.\ S.\ Plyushchay, R.\ Sasaki, and an anonymous referee are also acknowledged. \par
%
%

\end{document}